\documentclass{aa}  

\usepackage{graphicx}
\usepackage{siunitx}
\DeclareSIUnit\msol{M_\odot}
\DeclareSIUnit\year{yr}
\usepackage{txfonts}
\usepackage{xcolor}
\usepackage{lipsum}
\usepackage{subcaption}         
\usepackage{lscape}             
\usepackage{placeins}           
                                
\usepackage[colorlinks=true, linkcolor=blue, citecolor=blue, filecolor=blue, urlcolor=blue]{hyperref}
\newcommand{\eq}[1]{Equation \ref{#1}\xspace}

\newcommand{\fig}[1]{Fig. \ref{#1}\xspace}
\newcommand{\figstart}[1]{Figure \ref{#1}\xspace}
\newcommand{\sect}[1]{Section \ref{#1}\xspace}
\newcommand{\ardent}{{\tt ARDENT}\xspace}

\newcommand{\edit}[1]{#1}

\begin{document}

   \title{Completeness and dynamical stability in exoplanet occurrence rates}

   \subtitle{}

%

   \author{T. A. Baycroft\inst{1}\fnmsep\thanks{Corresponding author: \email{tbaycroft@sjtu.edu.cn}}
        \and M. Stalport \inst{2, 3}
        \and R. Eltschinger \inst{4, 5, 6}
        \and M. Pinamonti \inst{7}
        \and J. Davoult \inst{8}
        \and K. Al Moulla \inst{9}
        \and Y. Alibert \inst{4, 5, 6}
        }

   \institute{Tsung-Dao Lee Institute, Shanghai Jiao Tong University, 1 Lisuo Road, Shanghai 201210, China
   \and Space Sciences, Technologies and Astrophysics Research (STAR) Institute, Université de Liège, Allée du 6 Août 19C, 4000 Liège, Belgium
   \and Astrobiology Research Unit, Université de Liège, Allée du 6 Août 19C, 4000 Liège, Belgium
   \and Space Research \& Planetary Sciences (WP), Universität Bern, Gesellschaftsstrasse 6, 3012 Bern, Switzerland
   \and National Centre of Competence in Research - PlanetS, Universität Bern, Gesellschaftstrasse 6, 3012 Bern, Switzerland
   \and Centre for Space and Habitability (CSH), Universität Bern, Gesellschaftsstrasse 6, 3012 Bern, Switzerland
   \and INAF, Astrophysical Observatory of Turin, Via Osservatorio 20, 10025, Pino Torinese, Italy
   \and Institut für Weltraumforschung (WR), Deutsches Zentrum für Luft- und Raumfahrt (DLR), Rutherfordstraße 2, 12489 Berlin, Germany
   \and Instituto de Astrofísica e Ciências do Espaço, Universidade do Porto, CAUP, Rua das Estrelas, 4150-762 Porto, Portugal
   }

   \date{Received 7 August 2026/Accepted August 2026}

 
  \abstract
   {Exoplanet population studies are often quantified through the ``occurrence rate'' of certain types of planet. This value can be calculated from a survey of a sample of stars by taking the ``detection rate" and correcting with a completeness factor. This completeness is calculated via detection limits on each individual star, and can be calculated either individually for each detected planet or averaged in a certain area of parameter space.}
   {We test whether either the average completeness or individual completeness performs better than the other. We also test whether including dynamical stability constraints in the detection limits improves the occurrence rate results.} 
   {We use synthetic sets of exoplanetary systems and calculate occurrence rates for each and compare to what the value would be with perfect information. We compare how close each method is to the ground truth to assess which is most suitable.}
   {Overall, in our synthetic sets, the average completeness is more reliable though it can be a problem if there is a sharp change in the underlying population within the area of parameter-space considered. We find that using dynamical stability conditions causes a bias in the occurrence rate and do not recommend using these.}
   {}

   \keywords{methods: statistical -- planets and satellites: dynamical evolution and stability -- planets and satellites: general}

   \maketitle
   \nolinenumbers

\section{Introduction}

With the number of exoplanets ever-growing \citep{christiansen_nasa_2025}, studies of the exoplanet population statistics, which began in the 2010s \citep[e.g.][]{howard_occurrence_2010,gould_frequency_2010,johnson_giant_2010,mayor_harps_2011}, are becoming more numerous. This is expected to increase further with large samples of exoplanets expecting to be detected by future space missions such as PLATO \citep{rauer_plato_2025}, \edit{Earth 2.0 (ET)} \citep{ge_et_2022}, and the \edit{\textit{Nancy Grace Roman}} Space Telescope \citep{spergel_wide-field_2015,penny_predictions_2019}, as well as the upcoming data releases of {\it Gaia} \citep{gaia_collaboration_gaia_2016,lammers_exoplanet_2026}.

One main value calculated in exoplanet population statistics is the planetary ``occurrence rate" or ``planet frequency", a measure of how common planets with certain parameters are. This can be calculated as a value with uncertainties in a region of interest or as a parameterised function of, for example, orbital period \citep[see e.g.][]{fulton_california_2021}. The occurrence rate can also be calculated as the number of planets per star \citep[sometimes referred to as the planet frequency;][]{mayor_harps_2011} or as the fraction of stars with at least one planet in the required range \citep[e.g.][]{montet_trends_2014}. In this work we calculate the value of occurrence rates in discrete regions of parameter-space (planet \edit{m}ass-\edit{p}eriod) using the number of planets per star. We expect the results to be relevant to the other cases as well.

To calculate the occurrence rate in a given region of parameter space, the number of planets detected (in that region) per star sampled is measured. This then needs to be corrected for the planets that have likely been missed, called the completeness correction. Completeness refers to the completeness of the survey; if the survey is sensitive to a set of planet parameters in 40\% of stars, its completeness is 40\%. This completeness can be calculated in different ways, either averaged across the region of interest \citep[e.g.][]{pinamonti_hades_2022,bryant_occurrence_2023}, or only calculated for each detected planet that is included in the occurrence rate sum \citep[e.g.][]{howard_occurrence_2010,mayor_harps_2011}.

Completeness of a survey is usually calculated based on ``detection limits", these are a measure of in what region of parameter space a planet would have been detected should it exist. Detection limits are traditionally calculated through series of injection-recovery tests \citep{wittenmyer_frequency_2011,fulton_california_2021}, however other methods, such as sampling the parameter space consistent with noise in the data, exist \citep[e.g.][]{standing_bebop_2022}. Completeness, at a location in parameter space, is then the fraction of stars in the sample for which the data is sensitive to planetary signals. 

Detection limits classically use information from the data only. After subtracting the best-fit model, injection-recovery tests are performed on the residual radial velocity (RV) timeseries, exploring a range of RV semi-amplitudes $K$ and orbital periods $P$. Populating the ($P$,$K$) space with many injection-recovery tests, it is possible to construct a robust limit curve above which a certain percentage of signals is detected (e.g. the 95$\%$ detection limits). However, additional constraints besides the data can be provided. As such, the recent code \ardent \citep{stalport_ardent_2025} verifies the dynamical plausibility of the injected planets \footnote{\url{https://github.com/manustalport/ardent}}. Based on a combination of analytical criteria and, if needed, N-body simulations coupled with a chaos indicator, \ardent explores orbital stability in the period-mass space supposing an additional injected planet into the system. As a consequence of this stability check, entire ranges of orbital period can be excluded due to their dynamical instability, pushing down the detection limits to zero. It is important to note that as working hypothesis, \ardent supposes the stability of the nominal system (i.e. composed of the detected planets only). 

\begin{figure}
    \centering
    \includegraphics[width=0.99\linewidth]{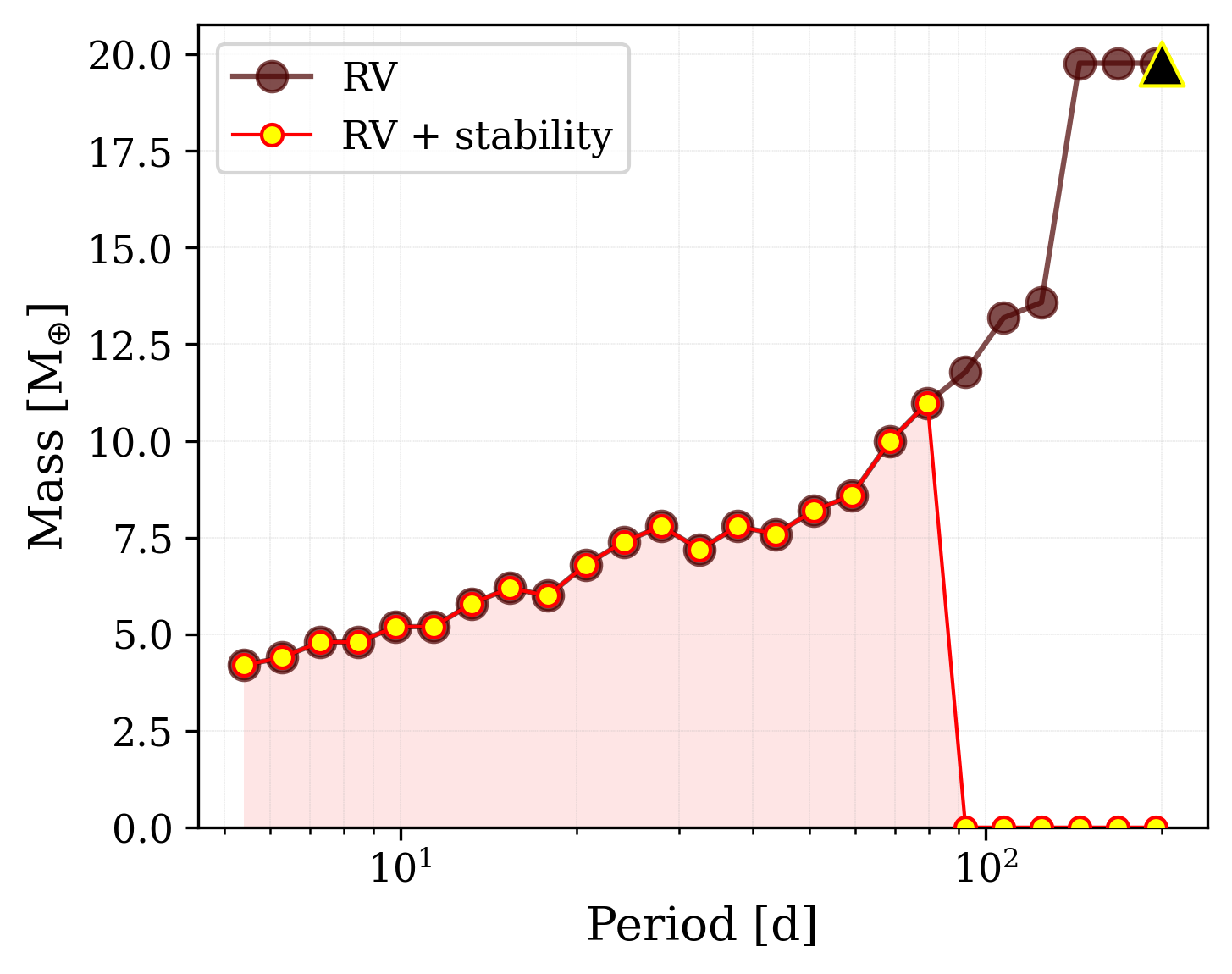}
    \caption{Illustrative \ardent run to compute planet detection limits in RV data. Based on a synthetic residual timeseries, we derived the 95$\%$ data-driven detection limits in brown. The dynamical detection limits are in red, computed using \ardent default settings and with 25 period bins. The black symbol represents the known planet, an eccentric ($e$=0.4) Jupiter-mass body (its mass is out of the range of the plot, hence we use an upward triangle).}
    \label{fig:ARDENT_illustration}
\end{figure}

\ardent computes detection limits both without and with the stability constraints. The former, based on the data only, are called ``data-driven" detection limits. The latter are referred to as ``dynamical" detection limits, due to their supplementary dynamical information \citep{stalport_ardent_2025}. We created a synthetic system to illustrate the use of \ardent. We supposed a 1-planet system composed of an eccentric giant planet ($e$=0.4, $P$=200d, $M$=1$M_J$), and created a residual RV timeseries composed of Gaussian white noise. We use a standard deviation of 2 m/s, and quote uncertainties as would originate from photon noise of 0.8 m/s. For this example, we generated 100 RVs spread over 300 days. We first derived the data-driven detection limits based on 5000 injection-recovery tests, as illustrated by the brown curve in \fig{fig:ARDENT_illustration}. Then, we computed the dynamical detection limits (in red-yellow). \ardent scans the period-mass space below the data-driven detection limits, and further rejects the dynamically unstable regions. In this example, the eccentric giant prevents the existence of any additional planet down to 90 days. \edit{If there are no detected planets, then a dynamical detection limit and a data-driven one are identical.}

The two methods represent different philosophies of what the detection limit is testing. Using data-driven limits is asking the question ``what planets would have been detected in the dataset we are using?" whereas the dynamical limits are asking ``what planets can we say do not exist in the systems we are observing?". It is unclear if one of these is better than the other; perhaps the dynamical limits are better since more information is being used. However, the use of detection limits with information from dynamical stability to calculate occurrence rates has not been tested.

In this work we investigate two aspects of exoplanet occurrence rate calculations, though these in principle can extend to any population inference. The first is comparing the average completeness to the individual completeness, assessing if one is typically more reliable than the other and the dependence on the underlying true population. The second is comparing the use of data-driven and dynamical detection limits.

One particular question that has been of interest is the occurrence rate of inner small planets conditional on the presence of an outer giant planet \citep[e.g.][]{bonomo_cold_2025,bryan_resolving_2025}. Since one of our investigations depends on orbital dynamics and stability, we chose to perform this study in the context of this particular setup. This choice is because the presence of an outer giant planet can have an important impact on the dynamics and stability of inner planetary systems \citep{ford_dynamical_2001,mustill_effects_2017,granados_contreras_dynamics_2018}. We expect that our results will nonetheless be applicable to all types of planetary systems. \edit{Similarly, while we apply this to radial velocity data the results should apply more broadly to other data types as well, whether using \ardent (once it has been expended to other datatypes) or using other software.}

In \sect{sec:methods} we describe the methods for generating synthetic planetary system samples and for calculating occurrence rates. In \sect{sec:results} we present results on the two aspects we investigate. We conclude in \sect{sec:conclusion}.

\section{Methods}\label{sec:methods}

\subsection{Synthetic planetary systems and data}

We use synthetic sets of planetary systems. We generate two simple sets, and use one more realistic set, a subset of the NG76longshot population \citep{emsenhuber_planetary_2023, emsenhuber_new_2025}.

To generate the simplistic sets of planetary systems a giant planet is chosen uniformly from the mass range \([120,620]{\,\rm M_\oplus}\) and period range \([600,1000]\) days. The number of inner planets is chosen randomly in the range \([0,4]\). The allowed period range for inner planets is \([2,200]\) days, this is divided into four bins of equal size in log-space. Each planet is assigned randomly to one of these bins (with replacement) the period is chosen uniformly within the bin. If two planets' period ratio is smaller than \(3/2\) then one is drawn again, this is to ensure that the planetary system will likely be dynamically stable. We also numerically validate stability of these systems (see \sect{sec:stab}). The mass of the inner planets is in the range \([1,20]{\,\rm M_\oplus}\). We simulate two cases, one where the masses are drawn uniformly from this range, and one log-uniformly.

The NG76longshot population was generated using the Generation III Bern model. The Bern model is a numerical model of planetary system formation and evolution designed to generate synthetic planetary populations that can be directly compared with observations. Comprehensive descriptions of the model are provided in the NGPPS series of papers \citep{emsenhuber_new_2021, emsenhuber_new_2021-1}. Earlier developments of the model are presented in \citet{alibert_models_2005, mordasini_extrasolar_2009,emsenhuber_new_2021}.

Planet formation is described within the core accretion paradigm, where planetary embryos grow through the accretion of solid material and gas in a protoplanetary disc. The model simultaneously follows the evolution of the disc and the embedded planets, while accounting for processes such as gas-driven orbital migration and gravitational interactions between planets through N-body calculations. The formation phase is simulated over 20 Myr, corresponding to the lifetime of the gaseous disc. Following disc dispersal, the model tracks the subsequent thermal and structural evolution of the planets over longer timescales, up to 10 Gyr.

The NG76longshot synthetic population is based on 1,000 planetary system formation simulations around \(1\,M_{\odot}\) stars. Each simulation begins with 100 planetary embryos and follows their growth and dynamical evolution over a formation timescale of \SI{100}{\mega\year} \citep{emsenhuber_planetary_2023, emsenhuber_new_2025}. 

Each of the simplistic sets contains 500 planetary systems. The NG76longshot set contains 125 planetary systems, these are the ones out of the original 1000 that contain a gas giant (\(1{\rm\,au} \leq a \leq 10{\rm\,au}\), \(0.3{\rm\,M_{Jup}} \leq M\leq20{\rm\,M_{Jup}}\)). We chose to analyse multiple different sets of planets so that we can investigate how the underlying planet population impacts the occurrence rate calculation, in particular the uniformity (in log-space) of planet masses.

\edit{For the purposes of testing the assumptions in the occurrence rate calculation it is not actually necessary to generate RV timeseries or signals, artificial data-driven detection limits could have been imposed. \ardent can work with either a RV timeseries as its input, or a pre-generated detection limit. However to obtain the expected shape of a RV detection limit, we do generate RV timeseries with white Gaussian noise for each system and use \ardent to calculate the data-driven detection limits}. We generate a single sample of dates, covering 10 consecutive days out of every month for 8 months of each year for 10 years, with 20\% of data removed randomly based on expected bad weather or technical issues \citetext{Eltschinger et al. in prep}, corresponding to a very intense RV monitoring campaign. This is chosen to simulate the best-case observational scenario with lots of data, minimising aliasing and giving smooth detection limits. We generate integer JD values for the time and then add to each timestamp a random value drawn from \(\sim\mathcal{N}(0,0.1)\) days. We do not generate RV signals for the planets and simply generate Gaussian white noise with a range of amplitudes: \([1,5]\) m/s for the simplistic setups where the planet masses range from \([1,20]\,{\rm M_\oplus}\); and \([1,20]\) m/s for the NG76longshot setups where the planet masses range from \([1,100]\,{\rm M_\oplus}\). While we could simulate planetary signals and then retrieve them to better emulate the process of a RV survey, the aim here is to isolate the occurrence rate calculation itself so we can simplify the process. 

\subsection{Detection limits and occurrence rates}

\begin{figure}
    \centering
    \includegraphics[width=\columnwidth]{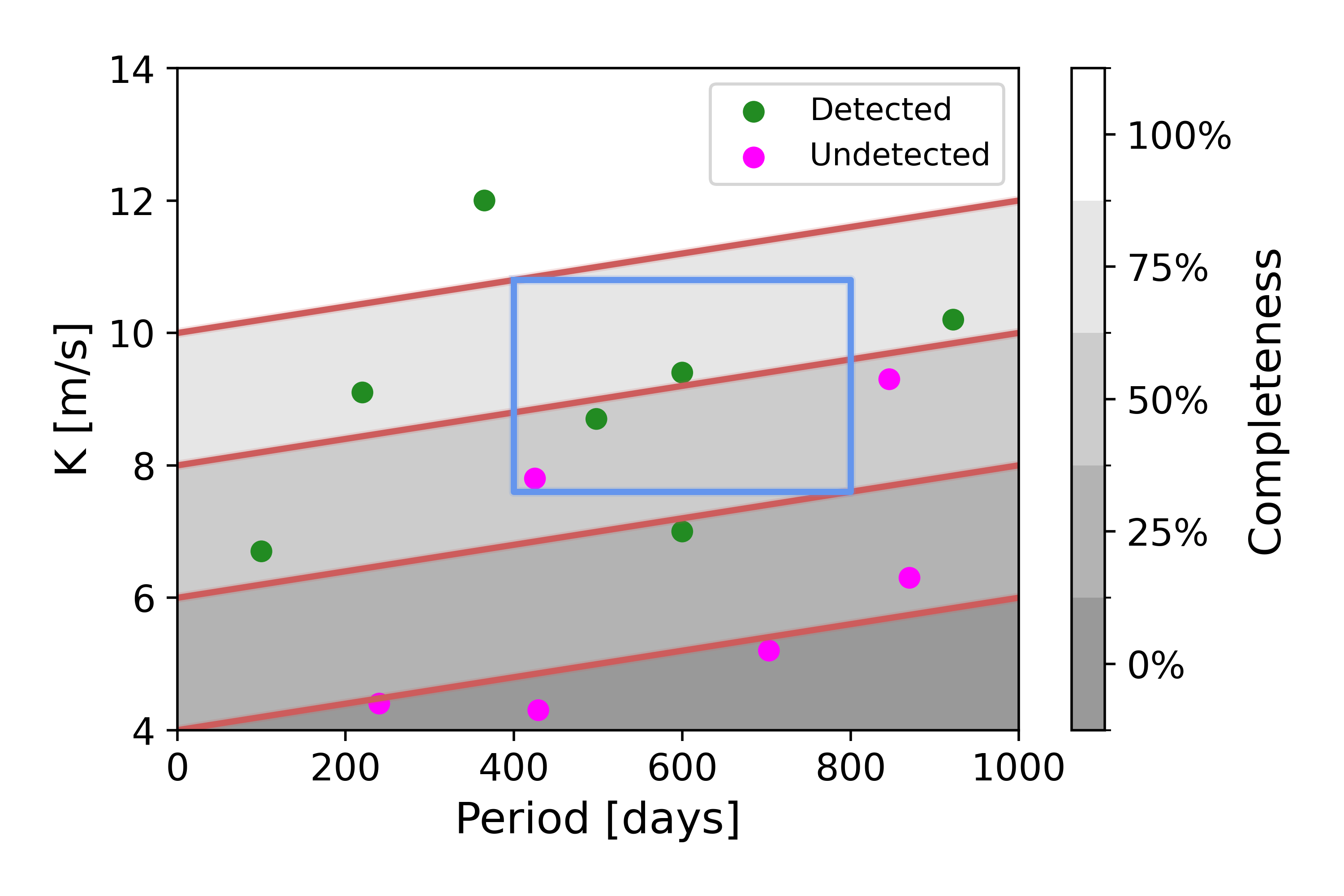}
    \caption{Illustration of completeness, four mock-detection limit lines \edit{in red} are shown along with a series of detected and undetected planets. The box shows the region discussed in the text.}
    \label{fig:comp_ex}
\end{figure}

We use \ardent to calculate detection limits from the white-noise time-series, these are the data-driven limits using only the information from the data. In each system if a planet\edit{'s RV semi-amplitude (calculated from the simulated parameters with all planets edge-on)} is above the detection limit we class it as ``detected", if below ``undetected". For each system, the data-driven detection limits and the detected planets are used within \ardent to calculate the dynamical detection limit.

We calculate occurrence rates in the two different ways introduced above, which we compare to each other. These ways differ essentially in how the completeness is calculated. The first, which we refer to as the ``individual completeness" method, uses the local completeness \(C_{\rm loc}\) at the location of each detected planet \citep[e.g.][]{mayor_harps_2011} to obtain the planet frequency or occurrence rate
\begin{equation}
    f_{\rm pl} = \frac{1}{N_\star}\sum_i\frac{1}{C_{\rm loc, i}},
\end{equation}
where \(N_\star\) is the total number of stars in the sample, and the sum is over all detected planets with \(C_{\rm loc, i}\) the local or individual completeness for each detected planet. \(C_{\rm loc, i} = N_{\rm DL,i}/N_\star\) where \(N_{\rm DL}\) is the number of detection limits in the sample that are below the \(i^{\rm th}\) detected planet , i.e. the number of stars in the sample around which this planet would have been detected. The planet frequency formula therefore simplifies to 
\begin{equation}
    f_{\rm pl} = \sum_i\frac{1}{N_{\rm DL, i}}.
\end{equation}
The uncertainty in \(\sigma_{f_{\rm pl}} = \frac{f_{\rm pl}}{\sqrt{N_\star}}\) following Poisson statistics \citep[e.g.][]{martin_bebop_2019}.

The second method, which we refer to as ``average completeness", takes the completeness averaged across the region of interest \(C_{\rm av}\) to obtain the effective number of stars \(n = N_\star C_{\rm av}\) in the region. This, along with the number of detected planets in the region \(k\) and a grid of planet frequencies \(f_{\rm pl}\), can be used to sample the probability density function \citep[e.g.][]{pinamonti_hades_2022}

\begin{equation}\label{eq:poisson}
    \mathcal{P}(k|n,f_{\rm pl}) = \frac{(nf_{\rm pl})^k e^{-nf_{\rm pl}}}{k!}.
\end{equation}
This allows to obtain the mode and the \(16^{\rm th}/84^{\rm th}\) percentiles of the probability density function.

The methods are illustrated in \fig{fig:comp_ex}, where a set of planets are shown along with four ``detection-limit" lines, giving regions of 0, 25\%, 50\%, 75\%, and 100\% completeness. Consider the region inside the box. With the individual completeness method the two detected planets would have completeness 75\% and 50\% respectively. With the average completeness method both would have 62.5\% completeness.

For each set of planets and each occurrence rate method, we calculate a ``ground-truth" occurrence rate, which represents what that sample would yield given perfect information. To do this we calculate the occurrence rates as above but use all of the generated planets rather than the detected ones, and we say that the completeness is 100\% everywhere. The value of occurrence rate and associated uncertainty then become the truth to which we compare the other calculations.

To use \eq{eq:poisson} to obtain the occurrence rate in a region of parameter space requires a single value of \(n\). Since \(n\) depends on the completeness in that region, using the Poisson pdf is not possible for the individual completeness method. This is why the two methods are implemented slightly differently. To check that this in not introducing a large difference we compare the occurrence rates calculated in the truth case with both methods. Both methods return the same values, and the width of the \(1\sigma\) confidence interval is the same to within 4\%. We deem this inconsistency small enough that it will not have much influence on the results.

 
\section{Results}\label{sec:results}

In each set of planetary systems we calculate the occurrence rate in 9 boxes of equal size in log-space. We judge the results from the different occurrence rate calculations by comparing the values in each box with the ground truth values. This is done by calculating the difference between the measured value and the truth weighted by the uncertainties added in quadrature to obtain
\begin{equation}
    \sigma_f = \frac{f_{\rm meas} - f_{\rm true}}{\sqrt{\sigma^2_{f_{\rm meas}} + \sigma^2_{f_{\rm true}}}},
\end{equation}
where \(\sigma_{f}\) is the measured weighted distance between a measured occurrence rate \(f_{\rm meas}\) and the true one \(f_{\rm true}\), with uncertainties in each of these \(\sigma_{f_{\rm meas}}\) and \(\sigma_{f_{\rm true}}\). The sign of this difference is retained such that knowledge of whether the value is overestimated or underestimated is maintained. The weighted distance from the truth is compared for the different methods through their difference \(\Delta = \sigma_{f,1} - \sigma_{f,2}\).

As mentioned above, the truth values for the two completeness methods are the same, so in the limit of perfect information they give the same result. The individual completeness method does not give a value of completeness if there are no detections, so the various results boxes that depend on this calculation are left blank.

\subsection{Different completeness methods}

\begin{figure*}
    \centering
    \textbf{Simplistic set - Uniform mass distribution}
    \includegraphics[width=\linewidth]{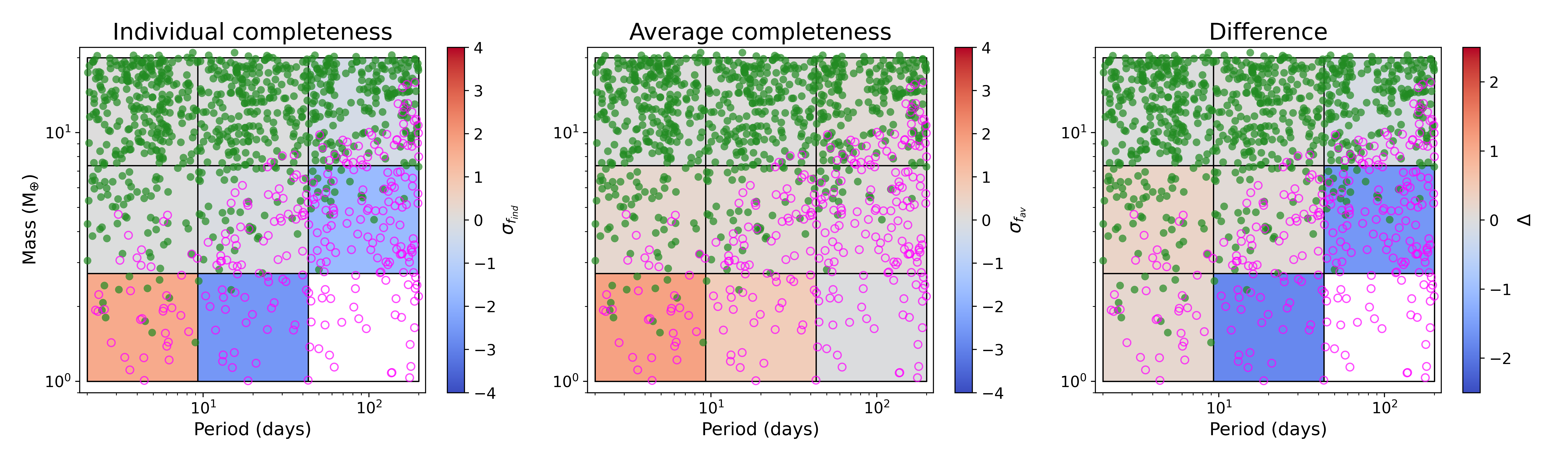}
    \textbf{Simplistic set - LogUniform mass distribution}
    \includegraphics[width=\linewidth]{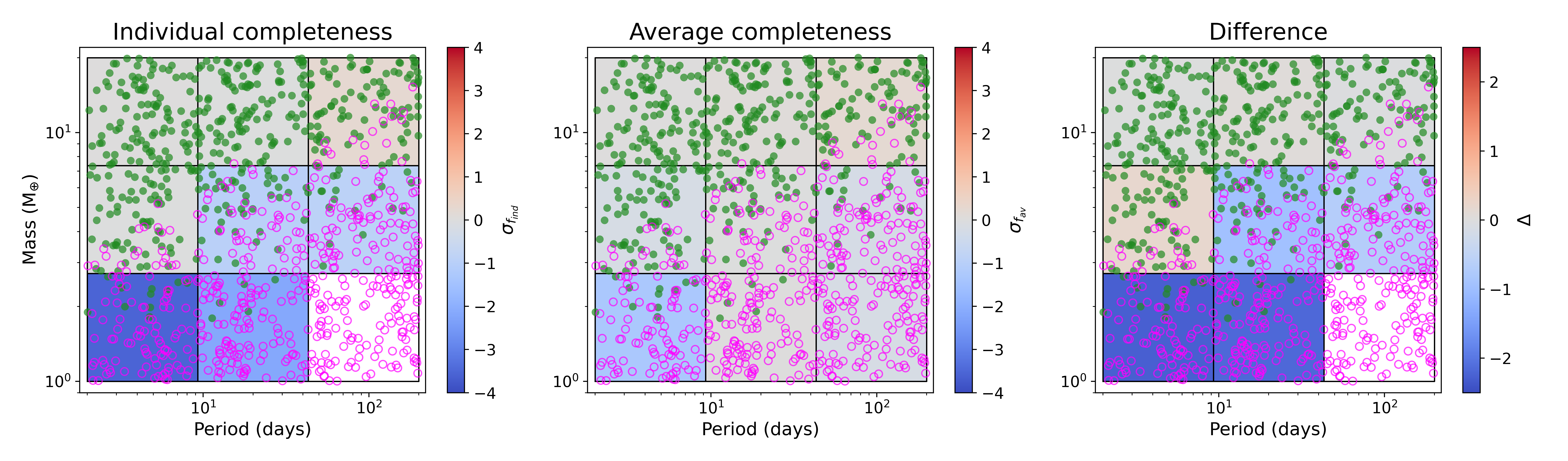}
    \textbf{NG76longshot set}
    \includegraphics[width=\linewidth]{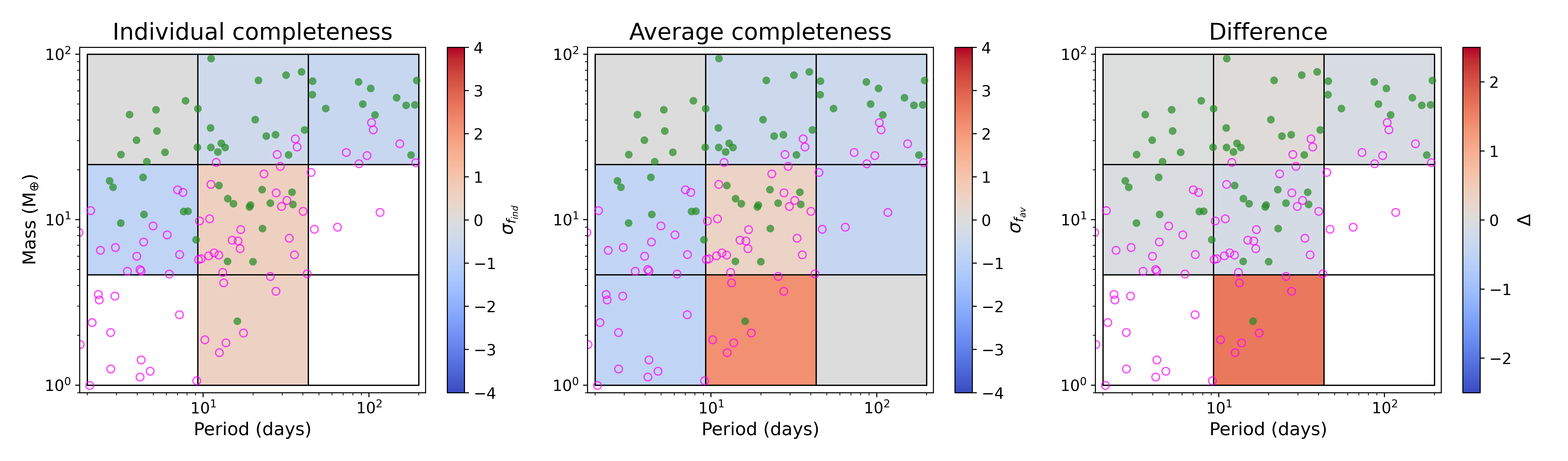}
    \caption{Comparison of methods for completeness calculation. Top row: simplistic set, uniform mass distribution. Middle row: simplistic set, log-uniform mass distribution. Bottom row: more realistic NG76longshot set. Left column: weighted difference between occurrence rate, calculated using the individual completeness method, and the truth. Blue indicates an underestimate and red an overestimate. Middle column: weighted difference between occurrence rate, calculated using the average completeness method, and the truth. Blue indicates an underestimate and red an overestimate. Right column: difference between the absolute values of  the two weighted differences. Here, blue means the average completeness is closer to the truth, and red that the individual completeness is closer to the truth. In all cases the set of planets detected/undetected are shown in green/magenta as filled/empty points.}
    \label{fig:methodcomp}
\end{figure*}

\figstart{fig:methodcomp} shows the results for all three synthetic samples of planetary systems comparing the two methods of calculating completeness. The set of planets (detected and undetected) are also shown. The left-hand column shows the weighted distances from the truth for the individual completeness method, the central column the same for the average completeness method. In these cases, red means the occurrence rate is overestimated and blue, underestimated. The final column is the difference between the absolute values of the other two such that red means that the average completeness performs worse, and blue that the individual completeness performs worse.

Both completeness methods have potential issues. The individual completeness method underestimates the occurrence rate if there are not detections spanning the whole range of interest. This is because information on the completeness is only included for each detected planet. This is especially present in the left panel of second row of \fig{fig:methodcomp}, in the lower regions there are many undetected planets at a lower mass than the smallest detected planet, information about these is unavailable to the individual completeness method, so the occurrence rate is drastically underestimated. In comparison for the same set of planets the average completeness does a much better job. This is because the average completeness inherently assumes that the distribution of planets is uniform throughout the box in which the occurrence rate is calculated, and this set of planets is the one with mass distributed log-uniformly (the completeness is calculated in log-space). In comparison the top row of \fig{fig:methodcomp} shows the results when the masses are not log-uniform, and the average completeness overestimates some of the occurrence rates. However even in this case it is performing better than the individual completeness.

In the potentially more realistic set of planets from NG76longshot, the two methods achieve the same answer in all boxes (likely due to the smaller number of systems) except one. This box is a clear case where assuming uniform planet distribution fails since there is a sharp cut-off halfway down the box, so the average completeness greatly overestimates the occurrence rate. Therefore, if the real exoplanetary populations contains similar sharp cut-offs, the average completeness method will be unsuitable for regions that cross this cut-off. 

Overall these results suggest that the average completeness is better. However, if there is a strong reason to believe the planet distribution is not uniform and planets are detected across the entire range of interest, then the individual completeness may be more appropriate.

\subsection{Using dynamical stability limits}\label{sec:stab}

\begin{figure*}
    \centering
    \textbf{Individual completeness}
    \includegraphics[width=\linewidth]{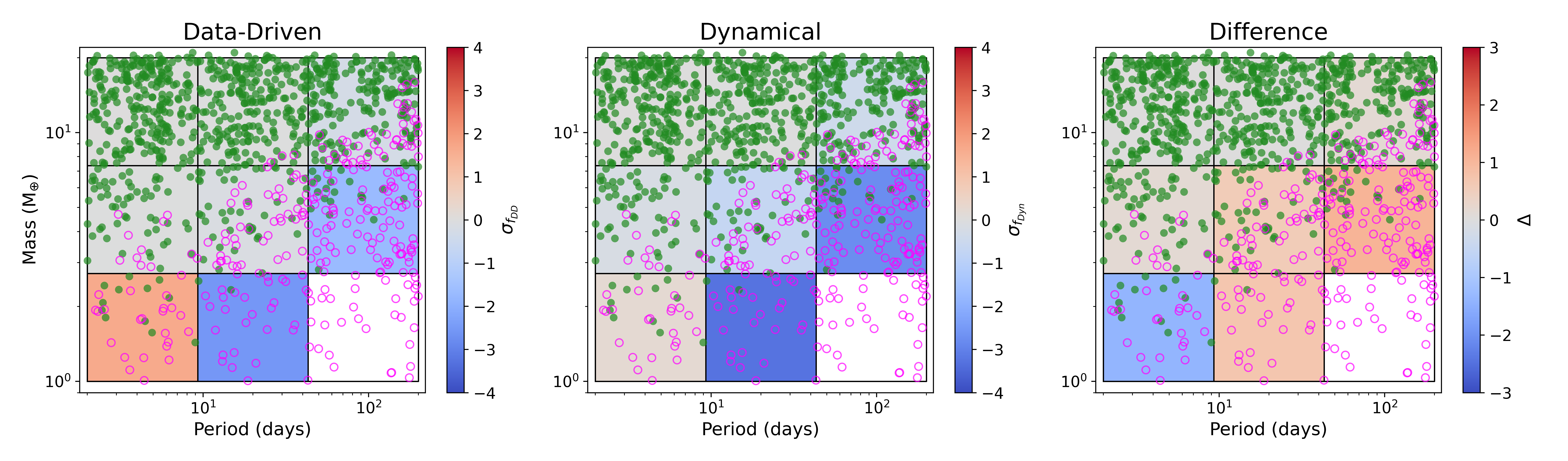}
    \textbf{Average completeness}
    \includegraphics[width=\linewidth]{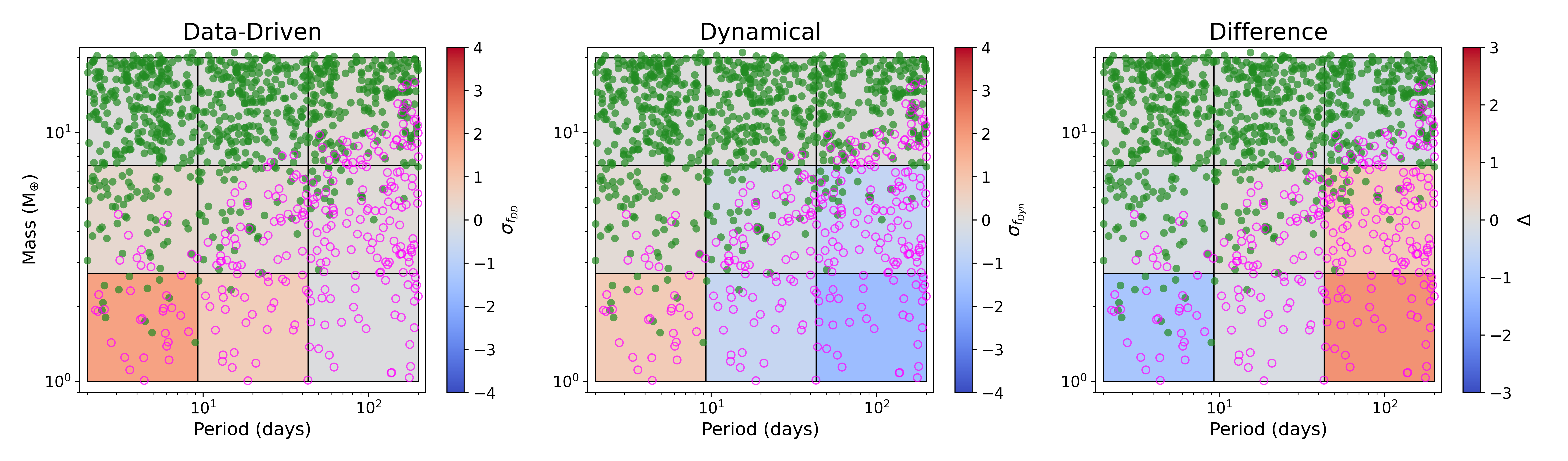}
    \caption{Comparison of methods for detection limit calculation, for the simplistic set of planetary systems with a uniform mass distribution. Top row: occurrence rate is calculated using the individual completeness. Bottom row: occurrence rate is calculated using the average completeness. Left column: weighted difference between occurrence rate, calculated using the data-driven detection limits, and the truth. Blue indicates an underestimate and red an overestimate. Middle column: weighted difference between occurrence rate, calculated using the dynamical detection limits, and the truth. Blue indicates an underestimate and red an overestimate. Right column: difference between the absolute values of the two weighted differences. Here, blue means the dynamical detection limit method is closer to the truth, and red that the data-driven detection limit method is closer to the truth. In all cases the set of planets detected/undetected are shown in green/magenta as filled/empty points.}
    \label{fig:dynTom2}
\end{figure*}

\begin{figure*}
    \centering
    \textbf{Individual completeness}
    \includegraphics[width=\linewidth]{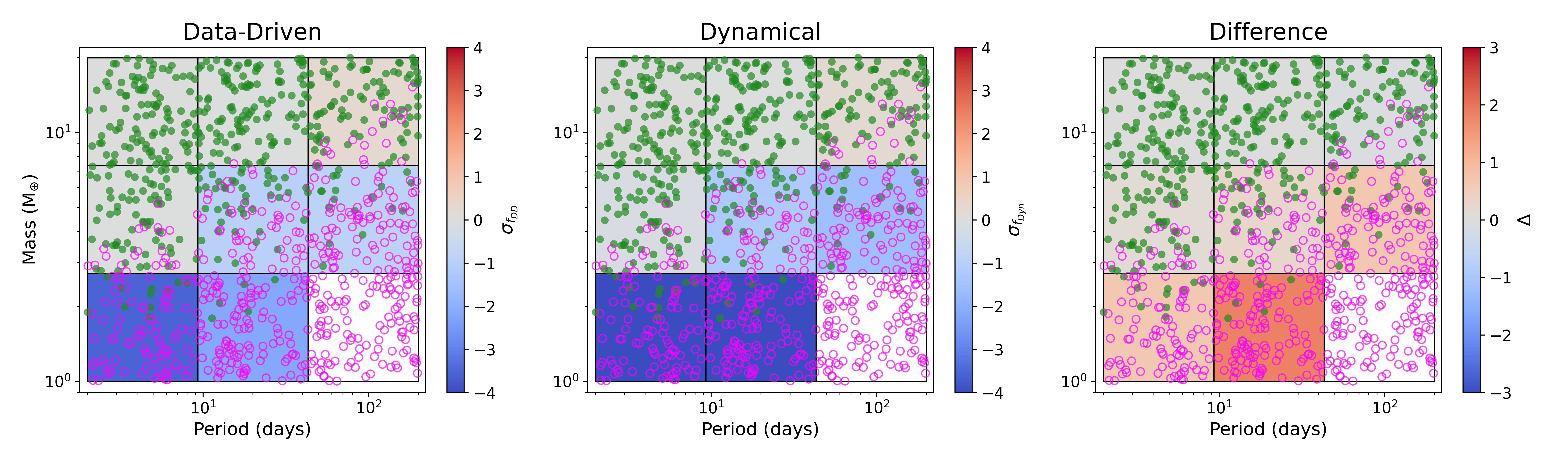}
    \textbf{Average completeness}
    \includegraphics[width=\linewidth]{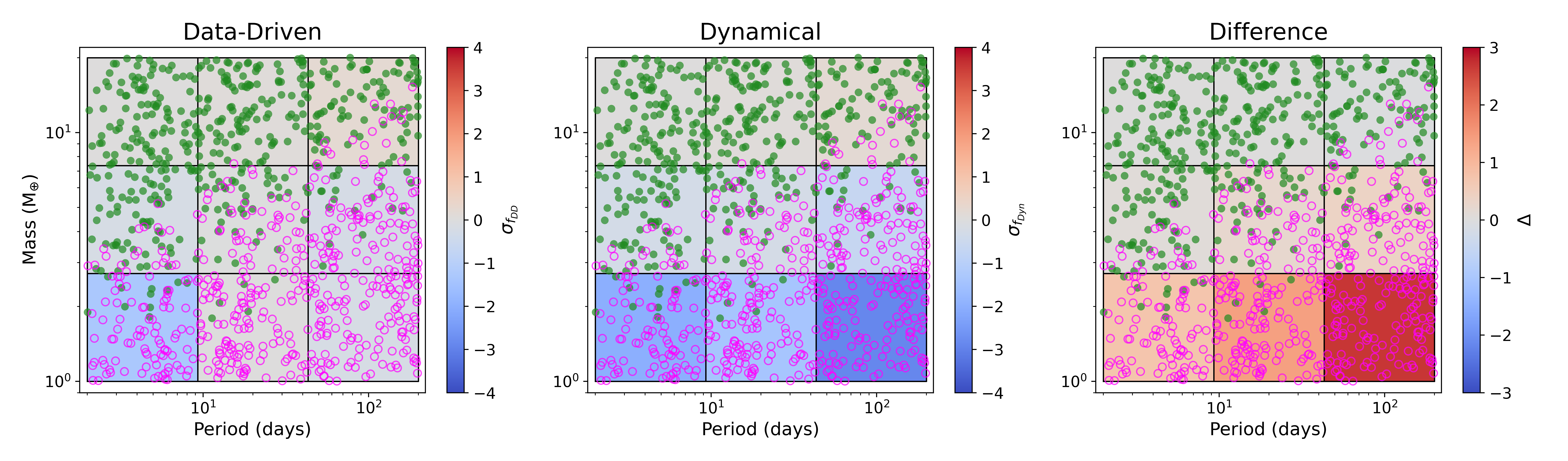}
    \caption{\edit{Same as \fig{fig:dynTom2} but} for the simplistic set of planetary systems with a log-uniform mass distribution.}
    \label{fig:dynTom3}
\end{figure*}

\edit{By restricting to systems that include gas giants, and simulating systems with multiple planets (sometimes multiple that are detected), we ensure that there will be a difference between the data-driven and dynamical limits. This will allow any potential biases to be clearer, even if in realistic scenarios it would be weaker.}

It is only appropriate to calculate the dynamical detection limit for a planetary system if the underlying system is stable. \ardent generate dynamical detection limits by testing stability when additional planets are included, but if the system would already be deemed unstable without additional planets, then the dynamical detection limit would take a value of 0 everywhere. Therefore, to ensure that it is appropriate to obtain dynamical detection limits for the synthetic systems, we test their long-term stability using the {\tt ARDENT\_TestStability} function. A system is integrated for up to \(10^6\) times the orbital period of the outermost body with a time-step 100 times smaller than the innermost orbital period. Instability is deemed if at some point as ``close encounter" occurs, defined as two bodies reaching within 3 Hill radii of each other; or if a body ``escapes", defined as reaching a radial distance greater than \(5\times\) the initial semi-major axis of the outermost body. In both of the simplistic sets all 500 pass this stability test so they are suitable to use for this purpose. In the NG76longshot set, however, \(\approx 1/3\) of the systems fail the stability test, most due to a close encounter. \citet{quintana_frequency_2016} \edit{determined that about} \(25\%\) of simulated systems \edit{with} super-earths have a giant impact after \(100 \,{\rm Myr}\). It thus makes sense that close encounters would still occur in some of these systems that were evolved for \(100 \,{\rm Myr}\), especially since they all have a gas giant which can increase the number of collisions in the inner part of the planetary system \citep{witt_outer_2024}. These late impacts mean that \ardent will not work for these systems. Removing all of the systems that fail this test from the NG76longshot set would leave too few for a useful analysis so we do not use this set of planets for the investigation of dynamical limits.

\figstart{fig:dynTom2} and \fig{fig:dynTom3} are similar to \fig{fig:methodcomp} in layout but now show the occurrence rates using the data-driven detection limits, the dynamical detection limits, and then the comparison between the two. This is done with both methods of completeness calculation for each of the two simplistic sets of planets. 

With the exception of one box where the dynamical limits perform better, all of the results point to either equivalent performance or the dynamical limits performing worse than the data-driven limits. They are typically equivalent in the larger-mass ranges where the completeness is high, and the difference gets larger where the completeness is lower. This makes intuitive sense, the dynamical detection limits can only be lower than the data-driven ones (see \fig{fig:ARDENT_illustration}), meaning that using them can only increase the completeness and therefore decrease the measured occurrence rate. In \fig{fig:dynTom2} and \fig{fig:dynTom3} the results from the dynamical method are always lower (more blue) than those from the data-driven method. Therefore, if there is no inherent bias in the data-driven method, \edit{and using the dynamical limits can only decrease the measured occurrence rate,} a bias must have been introduced by using the dynamical limits. 

\edit{There may indeed be biases in the data-driven method inherent to the calculation of occurrence rates by correcting for completeness based on detection efficiency. \citet{hsu_improving_2018} showed that for simulated \textit{Kepler} data this type of method did impose a bias where the occurrence rate was systematically underestimated. It is unclear if this bias would necessarily translate to radial velocities, though in many cases and especially for the individual completeness (\fig{fig:dynTom2} and \fig{fig:dynTom3}) we see that the occurrence rate is often underestimated. However in these cases, the dynamical detection limits underestimate the occurrence rate even more.}

Our interpretation of how this bias manifests is as follows. Consider the three boxes in the lower mass range in \fig{fig:dynTom3}. These boxes should have low completenesses, with few systems in the sample having data sensitive to such low-mass planets. Using the data-driven detection limits, the average completeness in each of these boxes is \([0.18,0.042,0.0021]\) (from smallest to highest period range, to 2 significant figures). Some systems with poor data quality such that no sensitivity at all is reached in the lowest mass box will have a detected planet that prohibits planets in all or some of the box. These could be larger terrestrial planets in the same mass range, or outer gas giants that impact stability in this region. When using the dynamical limits, these will contribute to the completeness in the lowest-mass box, since due to stability constraints the detection limit is placed at zero over a range near the planet's location. This leads to completenesses of \([0.22, 0.084, 0.030]\), which in one case is more than 10 times larger than before. 

\begin{figure}
    \centering
    \includegraphics[width=0.8\columnwidth]{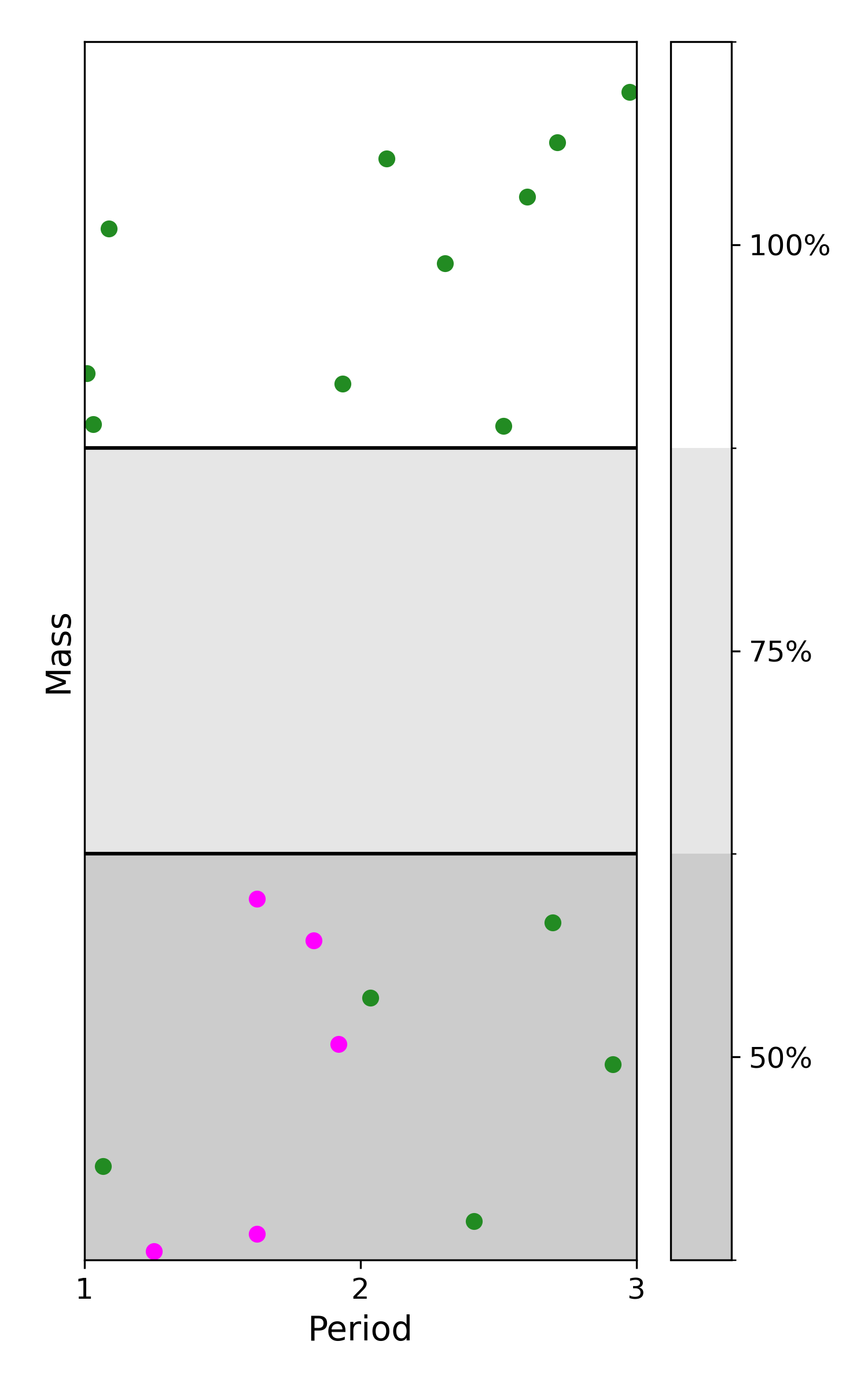}
    \caption{\edit{Demonstrative example of a sample of 20 stars with 10 hot-Jupiters (all detected) and 10 terrestrial planets (half of which are detected). The shades show the completeness of a hypothetical survey in each area.}}
    \label{fig:example}
\end{figure}

\edit{\figstart{fig:example} showcases a contrived example that demonstrates why the artificial increase in completeness from the use of dynamical detection limits leads to a bias. Consider a survey of 20 stars for planets in period range [1,3] days. Suppose that 10 of the stars have a hot-Jupiter, which will dynamically exclude a terrestrial planet in the period range, and the other 10 stars have a terrestrial planet. Suppose further a survey completeness of 50\% in the terrestrial-mass range and 100\% in the Jupiter-mass range, such that all the hot-Jupiters are detected and only 5 of the 10 terrestrials. This would be the case if half of the hot-Jupiter systems have detection limits curves below the terrestrial range, and half do not, and the same for the systems with terrestrial planets. Therefore \(1/4\) of the stars have a detected terrestrial planet, and \(1/2\) of the stars have a detected hot-Jupiter. Using the data-driven completeness of 50\% would recover the true occurrence rate of \(1/4 \div 1/2 = 1/2\), but using the dynamical detection limits, would mean that all the detection limits from hot-Jupiter stars are lowered to zero in this range, so the completeness would become 75\%, therefore the occurrence rate would be measured at \(1/4 \div 3/4 = 1/3\).}

\edit{Another} reason that the philosophy of ``what planets can we say do not exist in the systems we are observing?" fails is that in the method of measuring occurrence rates that we are using, there is an implicit assumption that a planet can exist anywhere within the region we consider with equal probability. This assumption is already not perfect since underlying exoplanet populations are likely not uniform, but it is even less applicable in the dynamical detection limit case. \edit{If the period range considered is wide enough, t}he existence of one (more massive) planet \edit{will} dictate what ranges of periods \edit{a second (less massive, i.e. below the data-driven detection limit) planet} can have \edit{within this range}. \edit{This does not mean that the probability that the smaller planet exists is exactly reduced by the proportion of area deemed unstable, which is what using the dynamical detection limits does. In the case of a narrow period range where all would be excluded, the issue mentioned in the example above still holds.} 

In one of the cases, the lowest mass and lowest period range of \fig{fig:dynTom2}, the data-driven occurrence rate is overestimated, likely an artifact of the relatively small number of planets in the box and many of them happening to have been detected. Therefore the dynamical occurrence rate is closer to the truth. However this is not correcting for the overestimation issue, but just happens to counter-balance it, only occurring in one case.

\edit{The differences between data-driven and dynamical rely on a range of different data-driven limit sensitivities. If all the stars in a sample have a very comparable data-driven limit then the results should not change if dynamical limits are used. This is because for a difference to ensue requires there to be planets that are detected in one system in an area that is below the data driven limit of a different system, but in a region that would be deemed unstable in that other system.}

Overall we would not recommend using dynamical detection limits to calculate occurrence rates within the current framework. If they are used in this way, care should be taken that the problem is posed in a way that is suitable. \edit{To incorporate dynamics unbiasedly, a new occurrence rate framework is needed.}

\section{Conclusion}\label{sec:conclusion}

We investigate exoplanet occurrence rates calculated from RV surveys with simulated planetary systems. We compare the use of different assumptions for completeness, average and individual. In the first, completeness is averaged over the range of interest, and in the second the completeness at the location of each detected planet is used. We find that the average completeness is in general more reliable, though if there are clear non-uniformities in the true population this can cause biases. We also compare the use of standard data-driven detection limits and dynamically informed detection limits in the occurrence rate calculation. We find that in general the use of orbital dynamics introduces a bias in the occurrence rate calculation resulting in an underestimated occurrence rate. This is particularly true in regions of low completeness. To include dynamical information into occurrence rates, a different or new framework should be developed tailored to its use. If such a framework can be created it could have powerful applications especially in regions where the completeness is low.  

\begin{acknowledgements}
Computations were performed using the TDLI-Astro cluster at Tsung-Dao Lee Institute.
KA acknowledges support from the Swiss National Science Foundation (SNSF) under the Postdoc Mobility grant P500PT\_230225.

\end{acknowledgements}

%
  \bibliographystyle{aa} 
  \bibliography{occurrence_rates_dynamics.bib} 

\begin{appendix}

\end{appendix}





\end{document}